\documentclass[journal]{IEEEtran}
\usepackage{float}
\usepackage{graphicx}
\usepackage{textcomp}
\usepackage{colortbl}
\usepackage{adjustbox}
\usepackage{caption}
\usepackage{paralist}
\usepackage{tabularx}
\usepackage{import}
\usepackage{algpseudocode,algorithm}
\usepackage{listings}
\usepackage{multirow}
\usepackage{caption}
\usepackage{subcaption}
\usepackage{hyperref}
\usepackage{circuitikz}
\usepackage{pgf}
\usepackage{pgfplots}
\usepackage{xcolor}
\usepackage{bm}
\usepackage{pgfplotstable}
\usepackage[many]{tcolorbox}
\pgfplotsset{compat=1.8}
\usepackage{adjustbox}
\usepackage{changepage}
\usepackage{enumitem}
\usepackage{dblfloatfix}
\usepackage{minted}
\usepackage{svg}
\usepackage{xpatch}

\usepackage[utf8]{inputenc}

\usepackage{amsmath}
\usepackage{amssymb}

\tcbset{
  highlightbox/.style={
    colback=gray!10,              
    colframe=gray!50,             
    boxrule=0.4pt,                
    arc=5pt,                      
    left=5pt,right=5pt,top=3pt,bottom=3pt, 
    enhanced,                     
  }
}

\usepackage{listings}
\usepackage{xcolor}

\lstdefinelanguage{Rust}{
  keywords={fn, let, mut, pub, where, Self, as, break, for, if, else, loop, return, impl, in, from},
  keywordstyle=\color{blue}\bfseries,
  ndkeywords={u8, usize, Scalar, ProjectivePoint, Choice, Double, PrimeField},
  ndkeywordstyle=\color{teal}\bfseries,
  identifierstyle=\color{black},
  sensitive=true,
  comment=[l]{//},
  morecomment=[s]{/*}{*/},
  commentstyle=\color{green!50!black}\ttfamily, 
  stringstyle=\color{orange}\ttfamily,
  morestring=[b]',
  morestring=[b]"
}

\definecolor{vscKeyword}{HTML}{0000FF}   
\definecolor{vscType}{HTML}{267F99}      
\definecolor{vscString}{HTML}{A31515}    
\definecolor{vscComment}{HTML}{008000}   
\definecolor{vscPreproc}{HTML}{AF00DB}   
\definecolor{vscFunc}{HTML}{795E26}      

\lstdefinestyle{vscodeLightC}{
  language=C,
  basicstyle=\ttfamily\footnotesize,
  numbers=none, numberstyle=\tiny\color{gray},
  frame=none, breaklines=true, columns=fullflexible, keepspaces=true,
  showstringspaces=false,
  commentstyle=\color{vscComment}\ttfamily,
  stringstyle=\color{vscString}\ttfamily,
  keywordstyle=[1]\color{vscKeyword}\bfseries,   
  keywordstyle=[2]\color{vscType}\bfseries,      
  keywordstyle=[3]\color{vscPreproc}\bfseries,   
  morekeywords=[1]{if,else,for,while,return,break,continue,sizeof},
  morekeywords=[2]{int,uint32_t,uint64_t,size_t,void,char,unsigned,signed,const,static,struct,union,enum,volatile},
  otherkeywords={\#define,\#include,\#if,\#ifdef,\#ifndef,\#endif,\#pragma},
  keywordstyle=\color{vscKeyword}\bfseries,
  emph={memset,p256_double,p256_add_mixed,CCOPY,NEQ,EQ},
  emphstyle=\color{vscFunc}
}

\lstdefinestyle{vscodeLightGo}{
  language=Go,
  basicstyle=\ttfamily\footnotesize,
  numbers=none,
  numberstyle=\tiny\color{gray},
  frame=none,
  breaklines=true,
  columns=fullflexible,
  keepspaces=true,
  showstringspaces=false,
  commentstyle=\color{vscComment}\ttfamily,
  stringstyle=\color{vscString}\ttfamily,
  keywordstyle=[1]\color{vscKeyword}\bfseries,   
  keywordstyle=[2]\color{vscType}\bfseries,      
  morekeywords=[1]{if,else,for,range,return,break,continue,go,defer,select,switch,case,default,func,var,const,type,package,import},
  morekeywords=[2]{bool,byte,rune,string,int,int8,int16,int32,int64,uint,uint8,uint16,uint32,uint64,uintptr,float32,float64,complex64,complex128,struct,interface,map,chan},
  emph={p256BaseMult,p256SelectAffine,p256PointAddAffineAsm,p256NegCond,p256MovCond,boothW6,NewP256Point},
  emphstyle=\color{vscFunc}
}

\newlength{\bubblesep}
\newlength{\bubblewidth}
\AtBeginDocument{\setlength{\bubblewidth}{.75\textwidth}}
\definecolor{bubblegreen}{RGB}{103,184,104}
\definecolor{bubblegray}{RGB}{241,240,240}

\newcommand{\bubble}[4]{%
  \tcbox[
    on line,
    arc=4.5mm,
    colback=#1,
    colframe=#1,
    tcbox width=auto limited, 
    width=\columnwidth, 
    #2,
  ]{\color{#3}#4}%
}

\ExplSyntaxOn
\seq_new:N \l__ooker_bubbles_seq
\tl_new:N \l__ooker_bubbles_first_tl
\tl_new:N \l__ooker_bubbles_last_tl

\NewEnviron{rightbubbles}
 {
  \begin{flushright}
  \sffamily
  \seq_set_split:NnV \l__ooker_bubbles_seq { \par } \BODY
  \int_compare:nTF { \seq_count:N \l__ooker_bubbles_seq < 2 }
   {
    \bubble{bubblegreen}{rounded~corners}{white}{\BODY}\par
   }
   {
    \seq_pop_left:NN \l__ooker_bubbles_seq \l__ooker_bubbles_first_tl
    \seq_pop_right:NN \l__ooker_bubbles_seq \l__ooker_bubbles_last_tl
    \bubble{bubblegreen}{sharp~corners=southeast}{white}{\l__ooker_bubbles_first_tl}
    \par\nointerlineskip
    \addvspace{\bubblesep}
    \seq_map_inline:Nn \l__ooker_bubbles_seq
     {
      \bubble{bubblegreen}{sharp~corners=east}{white}{##1}
      \par\nointerlineskip
      \addvspace{\bubblesep}
     }
    \bubble{bubblegreen}{sharp~corners=northeast}{white}{\l__ooker_bubbles_last_tl}
    \par
   }
   \end{flushright}
 }
\NewEnviron{leftbubbles}
 {
  \begin{flushleft}
  \sffamily
  \seq_set_split:NnV \l__ooker_bubbles_seq { \par } \BODY
  \int_compare:nTF { \seq_count:N \l__ooker_bubbles_seq < 2 }
   {
    \bubble{bubblegray}{rounded~corners}{black}{\BODY}\par
   }
   {
    \seq_pop_left:NN \l__ooker_bubbles_seq \l__ooker_bubbles_first_tl
    \seq_pop_right:NN \l__ooker_bubbles_seq \l__ooker_bubbles_last_tl
    \bubble{bubblegray}{sharp~corners=southwest}{black}{\l__ooker_bubbles_first_tl}
    \par\nointerlineskip
    \addvspace{\bubblesep}
    \seq_map_inline:Nn \l__ooker_bubbles_seq
     {
      \bubble{bubblegray}{sharp~corners=west}{black}{##1}
      \par\nointerlineskip
      \addvspace{\bubblesep}
     }
    \bubble{bubblegray}{sharp~corners=northwest}{black}{\l__ooker_bubbles_last_tl}\par
   }
  \end{flushleft}
 }
\ExplSyntaxOff

\ifCLASSOPTIONcompsoc
  \usepackage[nocompress]{cite}
\else
  \usepackage{cite}
\fi

\ifCLASSINFOpdf
\else
\fi
\begin{document}
%
\title{Sleep Reveals the Nonce: Breaking ECDSA using Sleep-Based Power Side-Channel Vulnerability}

\author{Sahan~Sanjaya,~\IEEEmembership{Student Member,~IEEE,} and~Prabhat~Mishra,~\IEEEmembership{Fellow,~IEEE,}

\IEEEcompsocitemizethanks{\IEEEcompsocthanksitem S. Sanjaya and P. Mishra are with the Computer \& Information Science \& Engineering, University of Florida, USA. Email: \{ssanjaya, prabhat\}@ufl.edu}
}

\markboth{}%
{Shell \MakeLowercase{\textit{et al.}}: A Sample Article Using IEEEtran.cls for IEEE Journals}


\maketitle

\begin{abstract}

Security of Elliptic Curve Digital Signature Algorithm (ECDSA) depends on the secrecy of the per-signature nonce. Even partial nonce leakage can expose the long-term private key through lattice-based cryptanalysis. In this paper, we introduce a previously unexplored power side-channel vulnerability that exploits sleep-induced power spikes to extract ECDSA nonces. Unlike conventional power-based side-channel attacks, this vulnerability leverages power fluctuations generated during processor context switches invoked by sleep functions. These fluctuations correlate with nonce-dependent operations in scalar multiplication, enabling nonce recovery even under constant-time and masked implementations.
We evaluate the attack across multiple cryptographic libraries, RustCrypto, BearSSL, and GoCrypto, and processor architectures, including ARM and RISC-V. Our experiments show that subtle variations in the power envelope during sleep-induced context switches provide sufficient leakage for practical ECDSA nonce extraction, recovering 20 bits of the nonce. 
These results establish sleep-induced power spikes as a practical cross-platform side-channel threat and highlight the need to reconsider design choices in cryptographic systems.

\end{abstract}


%
\IEEEpeerreviewmaketitle

\thispagestyle{empty}

\section{Introduction}
\label{sec:introduction}

Power side-channel attacks (PSCs) have emerged as one of the most powerful classes of physical attacks on cryptographic implementations. By monitoring the power consumption of a device, attackers can exploit data-dependent variations to recover sensitive information that remains secure at the algorithmic level. Since their introduction in the late 1990s, PSCs have evolved from targeting block ciphers such as DES, to symmetric primitives like AES, and eventually to asymmetric schemes including RSA and elliptic-curve cryptography. Despite decades of research and the deployment of countermeasures, PSCs continue to expose fundamental vulnerabilities in real-world systems.

Among asymmetric primitives, the Elliptic Curve Digital Signature Algorithm (ECDSA) has been a prominent target of side-channel research due to its reliance on a nonce during signature generation. Leakage of the nonce directly compromises the long-term signing key, and even partial information about the signing key can be exploited using lattice-based cryptanalysis. Over the years, numerous attacks have shown that biased, reused, or partially known nonces render ECDSA signatures insecure in practice. Deterministic ECDSA, as specified in RFC~6979~\cite{pornin2013deterministic}, was introduced to mitigate randomness failures by deriving nonces deterministically from the message hash and the secret key. However, this approach amplifies side-channel vulnerabilities, since repeated signatures on the same message produce identical nonces and thus identical leakage patterns.

Recently, the \texttt{SleepWalk}~\cite{sanjaya2025sleepwalk} side-channel vulnerability was identified, showing that a distinctive power spike appears during operating system context switches triggered by the \texttt{sleep} system call. Unlike traditional PSCs that rely on long, high-resolution traces and complex alignment, SleepWalk enables single-point leakage analysis based on the amplitude of the spike. Prior work demonstrated its feasibility on specific hardware platforms; however, the broader applicability of this vulnerability across architectures has not been fully explored.

In this paper, we show that the \texttt{SleepWalk} vulnerability is not confined to a single platform but can be observed across multiple processor architectures. Building on this insight, we demonstrate practical nonce-extraction attacks against ECDSA by exploiting \texttt{SleepWalk}-induced leakage. Specifically, we target three widely used cryptographic libraries: RustCrypto, BearSSL, and GoCrypto, running on two distinct processor architectures: ARM and RISC-V, and show that \texttt{SleepWalk} provides sufficient leakage to recover ECDSA nonces and compromise private signing keys. Specifically, this paper makes the following contributions:

\begin{compactitem}
   \item We propose a low-cost and faster attack on ECDSA. While prior attacks on ECDSA exist, they are often expensive. For example, Wang et al.~\cite{wang2023dvfs} require approximately one million data points, whereas our approach requires only one sample point per trace and at most 1000 traces.
   
   \item Prior efforts have primarily exploited vulnerabilities in constant-time scalar multiplication implementations. For instance,~\cite{wang2023dvfs} exploited scalar multiplication in the BearSSL library. In contrast, we demonstrate that the sleep-induced power spike vulnerability can be exploited across diverse scalar multiplication implementations, including RustCrypto, GoCrypto, and BearSSL.
   
   
   \item We establish that sleep-induced power spike vulnerability is a general and cross-architecture side-channel vulnerability that can be exploited to break ECDSA across both cryptographic libraries (RustCrypto, GoCrypto, BearSSL) and processor architectures (ARM, RISC-V).
\end{compactitem}

\vspace{0.05in}

To the best of our knowledge, \textit{this work is the first to show that the \texttt{sleep}-induced power spike is a general side-channel vulnerability enabling nonce-extraction attacks on three widely used cryptographic libraries.}

The remainder of this paper is organized as follows. Section~\ref{sec:background_and_related} provides background on ECDSA and surveys related efforts. Section~\ref{sec:sleepwalk_diff_arch} demonstrates the existence of \texttt{SleepWalk} across two different processor architectures. In Section~\ref{sec:ecdsa}, we reverse engineer the scalar multiplication implementations in three different software libraries and perform nonce-extraction attacks leveraging \texttt{SleepWalk}. Finally, Section~\ref{sec:conclusion} concludes the paper.

\section{Background and Related Work}\label{sec:background_and_related}

In this section, we first discuss the \texttt{SleepWalk} side-channel vulnerability and then ECDSA and its known nonce bits-based attacks. Next, we survey related efforts. Finally, we discuss the limitations of existing approaches to highlight the novelty of the proposed side-channel attack.


\subsection{SleepWalk Side Channel Vulnerability}
\label{subsec:sleepwalk}

\begin{figure}[t]
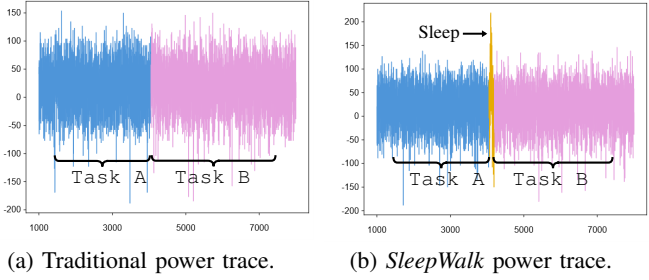

    \centering
    \begin{subfigure}{0.49\columnwidth}
        \centering
        \include{images/intro_traditional}
        \vspace{-0.41in}
        \caption{Traditional power trace.}
        \label{fig:traditional}
    \end{subfigure}
    \hfill
    \begin{subfigure}{0.49\columnwidth}
        \centering
         \include{images/intro-sleepwalk}
        \vspace{-0.33in}
        \caption{\textit{SleepWalk} power trace.}
        \label{fig:sleepwalk}
    \end{subfigure}
    \caption{Illustration of the \texttt{sleep}-induced power spike in ARM-based processor. Unlike traditional power analysis, this approach relies on a single power amplitude, thereby removing the need for full-trace analysis, advanced preprocessing, or external triggers~\cite{sanjaya2025sleepwalk}.}
    \label{fig:sleepwalk_traces}
\end{figure}

\texttt{SleepWalk} is a novel power side-channel vulnerability that exploits the distinctive power spike, as shown in Figure~\ref{fig:sleepwalk} produced during operating system context switches, which, in the approach in \cite{sanjaya2025sleepwalk}, were invoked through the kernel’s built-in \texttt{sleep} function. Unlike conventional side-channel analysis that depends on full power traces, advanced alignment, and external triggers, \texttt{SleepWalk} relies solely on the amplitude of this single, repeatable sampling point, thereby eliminating the need for high-end measurement equipment or complex preprocessing. The power spike itself is composed of two information-bearing components: a \textbf{context-switching power signature}, determined by the Hamming weight of register data at the moment of the switch, and a \textbf{residual power signature}, which reflects the lingering power footprint of the instructions executed just prior. The interplay between these components allows the spike amplitude to encode both instantaneous state and accumulated workload effects, making it possible to extract sensitive information with minimal overhead. As demonstrated through experiments in \cite{sanjaya2025sleepwalk}, this sleep-induced leakage channel significantly expands the attack surface by enabling adversaries to recover secrets from victim applications using only a single-point measurement, as opposed to traditional full-trace analysis.

\subsection{Elliptic Curve Digital Signature Algorithm (ECDSA)} \label{subsec:background_ecdsa}
 
ECDSA is a widely used public-key digital signature scheme defined over elliptic curve groups. Let $E$ be an elliptic curve defined over a finite field, with a base point $G$ of large prime order $n$. A signer holds a private key $p \in \mathbb{Z}_n$ and a corresponding public key $Q = [x]G$. To sign a message $m$, the signer first computes its hash $H(m) \in \mathbb{Z}_n$, then selects a uniformly random nonce $k \in \mathbb{Z}_n^{*}$. The signing procedure proceeds as follows: 
\begin{align}
    r &= ([k]G)_x \bmod n, \\
    s &= k^{-1}(H(m) + pr) \bmod n,
\end{align}
where $([k]G)_x$ denotes the $x$-coordinate of the point $[k]G$. The resulting signature is the pair $(r,s)$. 

The security of ECDSA critically depends on the secrecy and unpredictability of the nonce $k$, since exposure of $k$ immediately reveals the long-term signing key via $p = (sk - H(m))r^{-1} \bmod n$. Consequently, even partial leakage of $k$ across multiple signatures can enable powerful lattice-based key recovery attacks. In practice, failures in randomness generation have led to real-world compromises, most notably the recovery of Sony’s PlayStation~3 signing key, where predictable nonces rendered the private key recoverable~\cite{hotz2010console}.
 
To mitigate such randomness failures, RFC~6979~\cite{pornin2013deterministic} specifies a deterministic variant of ECDSA in which $k$ is derived deterministically as a pseudorandom function of the message hash and the secret key, commonly instantiated with Hash-based Message Authentication Code (HMAC). This design ensures that the same message always produces the same nonce, thereby eliminating dependence on external random number generators and preventing randomness failures.

\subsection{Known Nonce Bits-based Key Extraction on ECDSA} 
\label{subsec:background_ecdsa_hnp}

While deterministic ECDSA improves resilience against weak entropy sources, it introduces a new attack surface in the context of side channels: because identical messages always yield identical nonces, adversaries can repeatedly observe the same secret-dependent operations. This repetition amplifies side-channel leakage patterns, making side-channel assisted key-recovery attacks more practical.

When the nonce $k$ is not fully exposed, partial leakage of its bits can be sufficient to recover the secret signing key $p$. This is because ECDSA signatures $(r,s)$ reveal the linear relation 
\begin{align}
    sk \equiv H(m) + pr \pmod{n},
\end{align}
which directly connects the nonce $k$, the message hash $H(m)$, and the private key $p$. If an adversary can obtain several signatures for which some of the most significant or least significant bits of $k$ are known, these linear relations can be expressed as noisy modular equations~\cite{howgrave2001lattice}. Lattice-based techniques, such as those solving instances of the Hidden Number Problem (HNP)~\cite{boneh1996hardness}, can then be applied to efficiently recover $p$. Prior cryptanalytic results have shown that as few as 32-64 signatures with only a handful of known leading bits of $k$ are sufficient for full key recovery on common 256-bit curves~\cite{albrecht2021bounded, heninger2022rsa}. Consequently, side-channel attacks that leak even partial information about ECDSA nonces pose a severe practical threat, since they enable complete compromise of long-term signing keys from limited observations.


\subsection{Related Work}
\label{subsec:related}

We survey existing efforts that utilize side-channel analysis to extract information about the nonce used in ECDSA. 

\vspace{0.05in}
\noindent\textbf{Power Side Channel Attacks}:
PSC analysis exploits data-dependent variations in a device’s physical behavior, most commonly instantaneous power, to infer secret intermediates and keys. Classical techniques include non-profiled attacks like Simple Power Analysis (SPA)~\cite{kocher1996timing,ahmadi2023shield}, Differential Power Analysis (DPA)~\cite{kocher1999differential,aysu2018binary}, and Correlation Power Analysis (CPA)~\cite{brier2004correlation,benhadjyoussef2021power, sanjaya2025information} with Hamming-weight/distance leakage models, while profiled attacks learn device-specific leakage to reach near-optimal distinguishers~\cite{das2020electromagnetic, soroor2021deep}. Beyond cryptography, PSC has been applied to instruction-level disassembly~\cite{lipp2021platypus, glamovcanin2023instruction}, user-data inference~\cite{zhang2021red, wang2023dvfs}, and ML model extraction~\cite{wang2023powergan, xiang2020open}, aided by advanced statistical analysis and ML models. Modern works highlight that microarchitectural and power-management features (e.g., DVFS, RAPL) can enable remote or software-assisted variants and complicate defenses~\cite{gao2024deeptheft, lipp2021platypus, zhang2021red, wang2022hertzbleed, wang2023dvfs}.

\vspace{0.05in}
\noindent\textbf{Side-Channel Attacks on ECDSA}:
ECDSA’s security hinges on the per-signature nonce: any reuse, bias, or partial knowledge of the nonce enables recovery of the private key by casting the problem as a Hidden Number Problem (HNP) instance and solving it via lattice reduction~\cite{boneh1996hardness,nguyen2002insecurity,weiser2020big,albrecht2021bounded,breitner2019biased,de2013using,heninger2022rsa}. Implementation-level side channels magnify this risk: fixed/sliding-window scalar multiplication and table lookups can leak window digits, carries, or conditional additions through time, power, electromagnetic (EM), or cache footprints, enabling “known-nonce-bits” key recovery~\cite{moghimi2020tpm,genkin2016ecdsa,ryan2019return,jancar2020minerva,aranha2020ladderleak,ryan2019hardware,de2013using,belgarric2016side}. 

A scalar multiplication vulnerability in OpenSSL was exploited in~\cite{aranha2020ladderleak}. By applying a Flush+Reload cache-timing attack, the authors extracted information about the most significant bit of the nonce. Although the probability of recovery per trace was limited, the attack still solved the HNP with their proposed improvements, requiring between $2^{29}$ and $2^{35}$ signatures. In the case of Qualcomm’s hardware-backed Keystone Trusted Execution Environment (TEE),~\cite{ryan2019hardware} presented a microarchitectural attack that recovered an ECDSA private key. The cache-based technique leaked both the least-significant bit and the three most significant bits of the nonce. Another work,~\cite{ryan2019return}, described a simple cache attack on OpenSSL’s ECDSA implementation. A Flush+Reload strategy was used to determine the outcome of the modular reduction step.  

Timing leakage has been exploited by~\cite{moghimi2020tpm} to demonstrate that leading zero bits in the nonce correlated with the execution time of scalar multiplication inside a Trusted Platform Module (TPM). Work in~\cite{jancar2020minerva} highlighted a vulnerability across five cryptographic libraries on a smartcard chip. Their results revealed the bit length of the nonce by measuring the execution time of scalar multiplication.  

Electromagnetic (EM) leakage has also been leveraged for extracting information on ECDSA nonce. For example,~\cite{belgarric2016side} carried out an EM side-channel attack on Android smartphones, using Simple trace analysis to recover the number of least-significant zeroes in the nonce by observing the sequence of doublings before the last addition. Similarly,~\cite{genkin2016ecdsa} reported EM and power side-channel attacks against ECDSA implementations in OpenSSL, CoreBitcoin, and CommonCrypto running on mobile phones. They showed that scalar-dependent point multiplications could be identified from EM and power traces. In~\cite{de2013using}, a template-based power attack on smart cards exploited the modular inversion operation in ECDSA signing to recover low-order bits of the nonce. Since modular inversion is a variant of binary inversion whose execution time depends on the nonce and modulus, the authors were able to build templates for nonce recovery.  

Finally, empirical studies confirm that nonce-bit recovery attacks extend across commodity CPUs and embedded SoCs, and across multiple libraries using windowed scalar multiplication, highlighting that constant-time code alone is insufficient without leakage-resilient algorithms and system-level isolation~\cite{wang2023dvfs}.

\vspace{0.05in}
\noindent\textbf{Limitations of Existing ECDSA Side-Channel Attacks: }
Prior works on physical side-channel attacks against ECDSA have relied on vulnerabilities that do not appear in constant-time implementations~\cite{belgarric2016side, genkin2016ecdsa, de2013using}. Similarly, microarchitectural side-channel attacks have typically been limited to recovering only a few bits of the nonce~\cite{aranha2020ladderleak, ryan2019hardware, ryan2019return} or exploiting execution-time dependencies, which are absent in constant-time implementations~\cite{moghimi2020tpm, jancar2020minerva}. Attacks targeting constant-time ECDSA implementations~\cite{wang2023dvfs} have so far focused on a single library and were restricted to one hardware setup. In contrast, our work demonstrates a vulnerability that can be exploited to break ECDSA across multiple cryptographic libraries (RustCrypto, GoCrypto, BearSSL) and processor architectures (ARM, and RISC-V). Our experimental results further show that the attack can identify up to nine leading zero bytes in the nonce. Moreover, the power side-channel vulnerability exploited in our work requires only a minimal setup~\cite{sanjaya2025information}, in contrast to prior efforts that rely on privileged observability (e.g., advanced oscilloscopes, EM probes, power/RAPL-like interfaces, or microarchitectural performance signals), which do not consistently generalize across libraries and platforms.

\section{Cross-Architecture SleepWalk Vulnerability}
\label{sec:sleepwalk_diff_arch}


In this section, we evaluate the presence of the \texttt{SleepWalk} side-channel vulnerability across two different processor architectures: ARM, and RISC-V. In this experiment, we utilize the codes presented in~\cite{sanjaya2025sleepwalk} to generate two test scenarios: one corresponding to low residual power and the other to high residual power. For each test scenario, we vary the effect of Hamming weight (HWT) in registers during a context switch. These scenarios are then evaluated using the experimental setup described as follows.

\subsection{Experimental Setup}
\label{subsec:setup}

\begin{table}[h]
  \centering
  \caption{Single-board computers used in experiments.}
  \label{tab:eval_setup_sum}
  \begin{tabular}{| l | l | l |}
    \hline
    \textbf{SBC} & \textbf{Processor} & \textbf{Operating System} \\
    \hline
    Raspberry Pi~4B &
    ARM Cortex-A72 CPU &
    GNU/Linux Debian~12 \\
    \hline
    VisionFive~2 &
    RISC-V U74 CPU &
    GNU/Linux Debian~12 \\
    \hline
  \end{tabular}
  \vspace{-0.1in}
\end{table}

\begin{figure*}[t]
    \centering
    \scriptsize
    \begin{subfigure}{0.48\textwidth}
        \centering
        \vspace{-0.04in}
        \includegraphics[width=\textwidth]{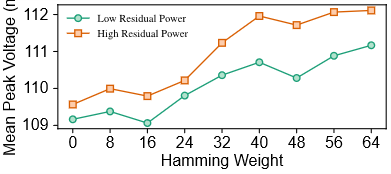}
        \caption{ARM architecture~\cite{sanjaya2025sleepwalk}.}
        \label{fig:sleepwalk_arm}
    \end{subfigure}
    \hfill
    \begin{subfigure}{0.48\textwidth}
        \centering
         \vspace{-0.1in}
         \includegraphics[width=\textwidth]{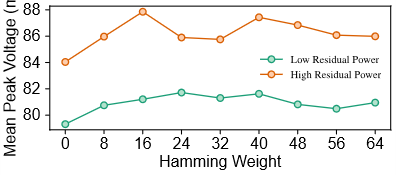}
        \caption{RISC-V architecture.}
        \label{fig:sleepwalk_risc_v}
    \end{subfigure}
    \caption{Effect of the \textbf{context-switching power signature} (green line) and the \textbf{residual power signature} (represented by the power difference between the green and orange lines).}
    \label{fig:sleepwalk_arch}
\end{figure*}

All the experiments in this paper are conducted on two different processor-architecture-based boards: a Raspberry Pi~4 Model~B single-board computer (SBC) with a BCM2711 System-on-Chip (SoC) integrating an ARM Cortex-A72 CPU, and a VisionFive~2 SBC with a JH7110 64-bit SoC featuring a RISC-V U74 quad-core CPU.

Details about the SBC and the operating system (OS) is presented in Table~\ref{tab:eval_setup_sum}. For power trace acquisition, a Keysight DSOX1102G oscilloscope~\cite{Keysight} was employed. The measurement probe was attached across the 5V supply and ground pins of the SBC board. Trace collection was automated via the Virtual Instrument Software Architecture (VISA) interface, enabling remote control of the oscilloscope. To reduce noise, a moving average filter with a window length of 10 samples was applied prior to peak detection. Python scripts handled both the automated capture process and the extraction of peak values, which were then archived for subsequent evaluation. These values were later utilized in different scenarios to demonstrate the existence of \texttt{SleepWalk} and to target partial nonce disclosure in ECDSA implementations across the RustCrypto, BearSSL, and GoCrypto libraries.


\subsection{Existence of SleepWalk in Different Processor Architectures}
\label{subsec:sleepwalk_diff_arch}

Figure~\ref{fig:sleepwalk_arch} depicts the results obtained from two different processor architectures for two test scenarios: one corresponding to low residual power and the other to high residual power. The green line represents the low residual power scenario and its variation with the HWT of the data in registers, while the orange line represents the high residual power scenario and its variation with HWT. The difference between the two lines indicates the presence of residual power effects in the \texttt{sleep}-induced power spike. Figure~\ref{fig:sleepwalk_arm} shows the results for the ARM architecture, which were validated in~\cite{sanjaya2025sleepwalk}. Figure~\ref{fig:sleepwalk_risc_v} shows the experimental results for the RISC-V architecture, demonstrating the existence of the \texttt{SleepWalk} vulnerability on this platform as well. We observe that the low residual power scenario exhibits a lower mean peak power compared to the high residual power scenario, and that the $0$-HWT case shows lower power consumption than non-zero HWT values. 

Therefore, an attack that differentiates between zero and non-zero values can exploit the \texttt{SleepWalk} vulnerability to perform a powerful side-channel attack. In the next section, we demonstrate how \texttt{SleepWalk} can be used to extract ECDSA nonces on three different software libraries across two different processor architectures.

\section{Breaking ECDSA using SleepWalk}
\label{sec:ecdsa}

In this section, we first describe the implementation of scalar multiplication in three software libraries, RustCrypto, Bear SSL, and GoCrypto. Next, we demonstrate nonce extraction attacks on these implementations.

\subsection{Scalar Multiplication Implementation} 
\label{sec:scalar_multiplication}

The core operation in ECDSA signing is scalar multiplication, where an elliptic-curve point $Q= [k]G$ is computed by repeatedly adding the base point $G$ to itself $k$ times. This operation dominates the computational cost of ECDSA and is a primary source of side-channel leakage. In practice, scalar multiplication is implemented using algorithms such as double-and-add, sliding-window, or fixed-window methods, which process the scalar $k$ bit by bit or bits chunk by chunk. The following subsections discuss scalar multiplication in RustCrypto, BearSSL, and GoCrypto libraries and explain how these constant-time implementations are vulnerable to power-side channel attacks.

\subsubsection{\textbf{RustCrypto}}

We first examine the scalar multiplication function used in RustCrypto ECDSA signing. In RustCrypto, ECDSA signing employs a deterministic ephemeral scalar $(k)$, where $k$ is the nonce computed using the algorithm described in RFC~6979. Consequently, exploiting bits of $k$ can lead to a key recovery attack, as explained in Section~\ref{subsec:background_ecdsa_hnp}. As shown in Listing~\ref{lst:mul_rust}, RustCrypto implements scalar multiplication using a constant-time fixed 4-bit window left-to-right double-and-add algorithm. This functionality can be found in $RustCrypto/elliptic-curves/primeorder/src/projective.rs$.

The function in Listing~\ref{lst:mul_rust} performs scalar multiplication, where it computes $[k]P$ for a given scalar $k$ and point $P$. When the argument $P$ corresponds to the base point $G$ of the elliptic curve, as in the case of ECDSA signing, it provides access to a precomputed lookup table $pc$ containing multiples of $G$: $[0]P, [1]P, \ldots, [15]P$ (lines 5-14). This 4-bit window table accelerates the computation by reducing the number of required point additions during the multiplication.

After the local lookup table $pc$ is computed, a loop (lines 17-29) is executed at an accumulator point $q$, which is initialized as the identity point (all-0 bit pattern) (line 15). The implementation selects $[slot]P$ from $pc$ (lines 20-22), and adds it into $q$ (line 23), where $slot$ is a four-bit nibble extracted from the scalar $k$ (line 18). For all but the final window, the code then applies four-point doublings to $q$ (line 27) and decrements $pos$ by $4$ (line 28). Therefore, an iteration of the main loop computes $q = [16](q + [slot]G)$.

The code uses several constant-time techniques and edge-case guards:
\begin{enumerate}
  \item {No data-dependent control flow.} The loop always starts at the most significant nibble boundary
  (${pos = ceil(NUM\_BITS/8)*8 - 4}$) and processes all windows. There is no early exit based on $k$.
  \item {Branch-free table lookup.} To obtain $[slot]G$, the code initializes $t$ to the identity and scans ${pc[1 \cdots 15]}$, using ${conditional\_assign}$ with a mask. When $slot = 0$, $t$ remains the identity ($t = 0$), so adding it is a no-operation, but the control flow and memory access pattern are unchanged.
  \item {Fixed amount of doubling.} Each non-final iteration performs exactly four doublings, independent of $slot$.
\end{enumerate}

\begin{listing}[t]  
\caption{Scalar multiplication function in RustCrypto}
\label{lst:mul_rust}
\begin{lstlisting}
fn mul(&self, k: &Scalar<C>) -> Self
where Self: Double,
{
  let k = Into::<C::Uint>::into(*k).to_le_byte_array();
  let mut pc = [Self::default(); 16];
  pc[0] = Self::IDENTITY;
  pc[1] = *self;
  for i in 2..16 {
    pc[i] = if i % 2 == 0 {
                Double::double(&pc[i / 2])
            } else {
                pc[i - 1].add(self)
            };
  }
  let mut q = Self::IDENTITY;
  let mut pos = (<Scalar<C> as PrimeField>::NUM_BITS.div_ceil(8) * 8) as usize - 4;
  loop {
   let slot = (k[pos >> 3] >> (pos & 7)) & 0xf;
   let mut t = ProjectivePoint::IDENTITY;
   for i in 1..16 {
       t.conditional_assign(&pc[i], Choice::from(((slot as usize ^ i).wrapping_sub(1)>>8) as u8 & 1),);
   }
   q = q.add(&t);
   if pos == 0 {
      break;
   }
   q = Double::double(&Double::double(&Double::double(&Double::double(&q))));
   pos -= 4;
  }
}
\end{lstlisting}
\vspace{0.15in} 
\end{listing}

After examining this implementation, we observe a potential vulnerability that can lead to side-channel attacks. Consider the following scenario. Suppose the nonce $k$ has one leading zero byte, represented as
\[
k = 0x00AB.
\]
According to the implementation, the intermediate value $q$ is initialized as the identity point (all-zero bit pattern). Then, \texttt{pos} is initialized to the most significant byte index, as shown in line 16,
\[
pos = 16 - 4 = 12.
\]
Next, as implemented in line 18, $slot$ is assigned the first four bits of the nonce $k$,
\[
slot = 0.
\]
Since the intermediate value $t$ is assigned as $[slot]G$, when $slot = 0$, $t$ remains the identity (all-zero bit pattern), as described in lines 19-22,
\[
t = all-zero.
\]
Therefore, $q$ is accumulated with the identity because of the value of $t$ (line 23),
\[
q = all-zero.
\]
When four-point doublings are performed, the projective point doubling function ensures that doubling an all-zero input produces an all-zero output, so $q$ remains the identity, as shown in line 27,
\[
q = all-zero.
\]
In the next iteration, in line 28, $pos$ is decremented by 4,
\[
pos = 8,
\]
which selects the next four bits of the nonce $k$,  yielding 0 (line 18),
\[
slot = 0.
\]
Thus, the previously discussed zero and identity propagation happen again. In the following iteration, $pos$ is reassigned to 4, which selects the third four-bit segment of $k$ as $slot$ (line 18),
\[
slot = A.
\]
Based on the value in $slot$, other variables are calculated and reassigned.  

However, we can observe that when the nonce $k$ begins with a sequence of 4-bit all-zero segments (leading all-zero nibbles) before the first non-zero 4-bit segment, the main loop in Listing~\ref{lst:mul_rust} processes the elliptic curve identity value (all-zero bit pattern). As a result, these initial loop iterations that handle all-zero nibbles exhibit a lower HWT and Hamming distance (HD) compared to iterations that process non-zero nibbles. This variation in HWT and HD directly affects the device's power consumption pattern, thereby leaking side-channel information about the number of leading zero nibbles in the nonce $k$. Moreover, while the all-zero nibbles are processed in the loop, most of the processor registers are also assigned all-zero bit patterns and zeros. This results in a residual power difference when performing scalar multiplication in Listing~\ref{lst:mul_rust}, where nonces with fewer leading all-zero nibbles exhibit higher residual power, while nonces with more leading all-zero nibbles exhibit lower residual power. This behavior can be exploited as part of the \texttt{SleepWalk} vulnerability, as shown in Section~\ref{sec:exp_results}.

\subsubsection{\textbf{BearSSL}}

ECDSA implementation in BearSSL's 64-bit platform on the NIST P-256 curve employs a constant-time fixed 4-bit window left-to-right double-and-add algorithm, similar to the implementation in RustCrypto. The functionality resides in \texttt{point\_mul\_inner} in \texttt{BearSSL/src/ec/ec\_p256\_m64.c} (Appendix Listing~\ref{lst:mul_bearssl}).

The function \texttt{point\_mul\_inner} takes as input a scalar $k$ and a precomputed window of affine points $\{[1]P,[2]P,\ldots,[15]P\}$. It initializes the accumulator point $Q$ in Jacobian coordinates as the identity, with a flag variable $qz$ used to indicate whether $Q$ is still the all-zero representation. This ensures that the first non-zero nibble of $k$ initializes $Q$ correctly without requiring a secret-dependent branch.

The main loop processes the scalar $k$ byte by byte, extracting two 4-bit windows per byte. For each nibble, the function applies four-point doublings to $Q$ (i.e., $Q \leftarrow [16]Q$), then extracts the next window value $bits$ from $k$. If $bits$ is non-zero, it selects the corresponding affine point $[bits]P$ from the precomputed window. This lookup is performed in constant time by iterating through all candidates in $W$ and using bitwise masking to copy only the matching entry.

Once the candidate point is retrieved, it is added to $Q$ using mixed Jacobian-affine addition. If $Q$ was still the identity (as tracked by $qz$), a conditional move copies the selected point directly into $Q$. Otherwise, a constant-time conditional copy merges the updated accumulator with the addition result. After each iteration, $qz$ is updated to reflect whether $Q$ has been initialized with a non-zero point. In summary, each loop iteration computes $Q \;\leftarrow\; [16]Q + [bits]P$, where $bits$ is the current 4-bit nibble of the scalar.

After reverse engineering the BearSSL implementation, we identified a similar vulnerability as in RustCrypto: nonces with leading zero 4-bit nibbles produce sequences of computations involving zero values. In this case, $Q$ is initialized with an all-zero bit pattern, which sets $bz = 1$. The variable $bits$ remains zero, and the point $T$ is also computed as an all-zero bit pattern. This behavior persists until all leading zero nibbles in $k$ are processed. As a result, the computations exhibit low HW/HD activity, leading to lower residual power consumption compared to non-zero scenarios, which can be exploited via \texttt{SleepWalk} vulnerability, as shown in Section~\ref{sec:exp_results}.

\subsubsection{\textbf{GoCrypto}}
\label{subsec:scale_mul_gocrypto}

We finally examine the scalar multiplication routine used in the GoCrypto P-256 implementation (Appendix Listing~\ref{lst:mul_go}). In GoCrypto, for ECDSA signing, the user can choose whether the nonce is deterministic (per RFC~6979) or not by setting an input argument in the sign function. The Go routine computes $[k]G$ using a fixed 6-bit signed-window method with Booth recoding and precomputed base-point tables. This constant-time algorithm processes the scalar $k$ from right-to-left and can be found in $go/src/crypto/internal/fips140/nistec/p256\_asm.go$.

The function $p256BaseMult$ performs scalar multiplication of the fixed base point $G$. It first extracts the initial 6-bit window from the scalar and applies Booth’s recoding ($boothW6$), which maps each window to a signed digit in $\{\pm 1, \ldots, \pm 32\}$ along with a sign bit. Using this digit, it selects the corresponding precomputed affine multiple from $p256Precomputed$ in constant time ($p256SelectAffine$), and initializes the accumulator point $p$ with that multiple. Subsequent iterations advance the bit index by six, extract the next 6-bit chunk, and again apply Booth recoding to obtain a signed digit. For each window, the code selects the appropriate affine point from the precomputed table and invokes the constant-time mixed-add routine $p256PointAddAffineAsm$. 

We observe that trailing 6-bit zero chunks in the nonce create sequences of zero-value dominant computations, resulting in lower power consumption compared to non-zero chunks. Similar to RustCrypto and BearSSL, GoCrypto also exhibits the potential vulnerability of leaking the number of zero bits in the nonce used in ECDSA. 

In the next section, we leverage this vulnerability to perform a power side-channel attack by incorporating the \texttt{SleepWalk} power side-channel vulnerability.

\subsection{Experimental Validation} 
\label{sec:exp_results}

In this section, we perform ECDSA nonce exploitation attacks on three different software implementations across two processor architectures. Specifically, we target the RustCrypto, BearSSL, and GoCrypto libraries by exploiting the vulnerabilities in their scalar multiplication routines, as explained in Section~\ref{sec:scalar_multiplication}. These libraries are executed on the experimental setup described in Section~\ref{subsec:setup}, resulting in six different configurations.

\begin{listing}[h]  
\caption{Attack execution flow}
\label{lst:exp_process}
\begin{lstlisting}
ECDSA Signing Initialization
loop in NUMBER_OF_TRACES {
    loop in NUMBER_OF_ITERATIONS{
        ECDSA Sign
    }
    Sleep
}
\end{lstlisting}
\vspace{0.2in} 
\end{listing}

Since the \texttt{SleepWalk} vulnerability captures the effects of both context-switching power and residual power, we implement an execution flow to amplify the residual power effect. As shown in Listing~\ref{lst:exp_process}, in each experimental configuration, after initializing the ECDSA signing process, a few iterations of the same signing operation (with the same message) are performed before capturing the \texttt{sleep}-induced power spike. This repetition increases the influence of residual power. Table~\ref{tab:config_sum} summarizes the number of traces ($ \# Traces$) and the number of loop iterations ($ \# Iterations$) used in our experiments.

\begin{table}[h]
\centering
\caption{Results for different hardware boards and software configurations.}
\label{tab:config_sum}
\begin{tabular}{|l|c|c|c|}
\hline
\textbf{Hardware Board} & \textbf{Library} & \textbf{\# Traces} & \textbf{\# Iterations} \\
\hline
                & RustCrypto & 1000 & 20 \\
Raspberry Pi~4B & BearSSL & 1000 & 750 \\
                & GoCrypto & 1000 & 750 \\
\hline
                & RustCrypto & 500 & 20 \\
VisionFive~2    & BearSSL & 500 & 250 \\
                & GoCrypto & 500 & 1000 \\
\hline
\end{tabular}
\vspace{-0.1in}
\end{table}

During these experiments, a randomly generated secret key is used as the private key $p$. For each test scenario, whether involving leading or trailing zero-bit chunks, or leading or trailing zero bits, four different random messages are selected that produce the same number of leading or trailing zeros in the nonce according to the scenario. Each data point shown in the results is obtained by averaging the mean peak power values of the four messages.

\subsubsection{RustCrypto}

\begin{figure}[h]
    \centering
    \scriptsize
    \begin{subfigure}{0.48\linewidth}
        \centering
        \vspace{-0.04in}
        \includegraphics[width=1\linewidth]{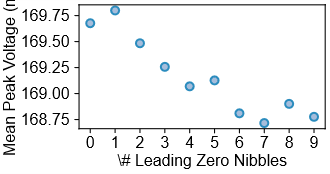}
        \caption{ARM}
        \label{fig:rasp_rust_nibbles}
    \end{subfigure}
    \hfill
    \begin{subfigure}{0.48\linewidth}
        \centering
         \vspace{-0.1in}
         \includegraphics[width=1\linewidth]{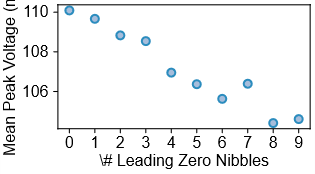}
        \caption{RISC-V}
        \label{fig:vision_rust_nibbles}
    \end{subfigure}
    \caption{Sleep-induced power spike vs number of leading zero \textbf{nibbles} in the nonce $k$ using \textbf{RustCrypto} ECDSA.}
    \vspace{-0.1in}
    \label{fig:rust_nibbles}
\end{figure}

\begin{figure}[h]
    \centering
    \scriptsize
    \begin{subfigure}{0.48\linewidth}
        \centering
        \vspace{-0.04in}
        \includegraphics[width=1\linewidth]{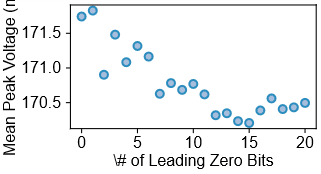}
        \caption{ARM}
        \label{fig:rasp_rust_bits}
    \end{subfigure}
    \hfill
    \begin{subfigure}{0.48\linewidth}
        \centering
         \vspace{-0.1in}
         \includegraphics[width=1\linewidth]{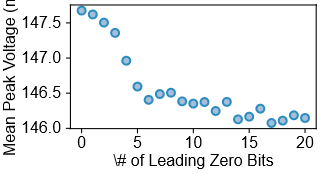}
        \caption{RISC-V}
        \label{fig:vision_rust_bits}
    \end{subfigure}
    \caption{Sleep-induced power spike vs number of leading zero \textbf{bits} in the nonce $k$ using \textbf{RustCrypto} ECDSA.}
    \vspace{-0.1in}
    \label{fig:rust_bits}
\end{figure}

Figures~\ref{fig:rust_nibbles} and~\ref{fig:rust_bits} depict the results obtained from experiments on the RustCrypto ECDSA implementation, following the execution sequence shown in Listing~\ref{lst:exp_process}. Figure~\ref{fig:rust_nibbles} shows that the mean peak power decreases with the number of leading zero nibbles, indicating that it is possible to distinguish non-zero leading-nibble nonces from those with leading-zero nibbles. Note that this power difference can be observed not only at the nibble level but also at the bit level, as illustrated in Figure~\ref{fig:rust_bits}.

\subsubsection{BearSSL}

\begin{figure}[h]
    \centering
    \scriptsize
    \begin{subfigure}{0.48\linewidth}
        \centering
        \vspace{-0.04in}
        \includegraphics[width=1\linewidth]{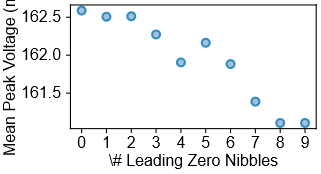}
        \caption{ARM}
        \label{fig:rasp_bearssl_nibbles}
    \end{subfigure}
    \hfill
    \begin{subfigure}{0.48\linewidth}
        \centering
         \vspace{-0.1in}
         \includegraphics[width=1\linewidth]{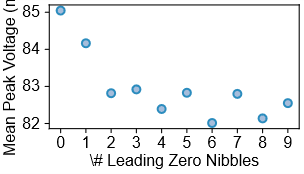}
        \caption{RISC-V}
        \label{fig:vision_bearssl_nibbles}
    \end{subfigure}
    \caption{Sleep-induced power spike vs number of leading zero \textbf{nibbles} in the nonce $k$ using \textbf{BearSSL} ECDSA.}
    \label{fig:bearssl_nibbles}
\end{figure}

\begin{figure}[h]
    \centering
    \scriptsize
    \begin{subfigure}{0.48\linewidth}
        \centering
        \vspace{-0.04in}
        \includegraphics[width=1\linewidth]{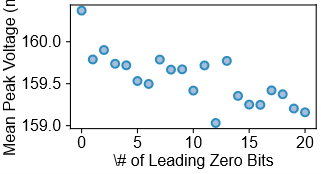}
        \caption{ARM}
        \label{fig:rasp_bearssl_bits}
    \end{subfigure}
    \hfill
    \begin{subfigure}{0.48\linewidth}
        \centering
         \vspace{-0.1in}
         \includegraphics[width=1\linewidth]{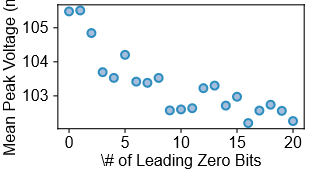}
        \caption{RISC-V}
        \label{fig:vision_bearssl_bits}
    \end{subfigure}
    \caption{Sleep-induced power spike vs number of leading zero \textbf{bits} in the nonce $k$ using \textbf{BearSSL} ECDSA.}
    \vspace{-0.1in}
    \label{fig:bearssl_bits}
\end{figure}

Similar to RustCrypto, the BearSSL implementation can also be exploited, as shown in Figures~\ref{fig:bearssl_nibbles} and~\ref{fig:bearssl_bits}, and the effect is observed irrespective of processor architecture. 

\subsubsection{GoCrypto}

\begin{figure}[h]
    \centering
    \scriptsize
    \begin{subfigure}{0.48\linewidth}
        \centering
        \vspace{-0.04in}
        \includegraphics[width=1\linewidth]{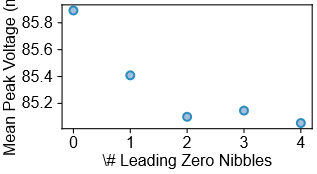}
        \caption{ARM}
        \label{fig:rasp_go_nibbles}
    \end{subfigure}
    \hfill
    \begin{subfigure}{0.48\linewidth}
        \centering
         \vspace{-0.1in}
         \includegraphics[width=1\linewidth]{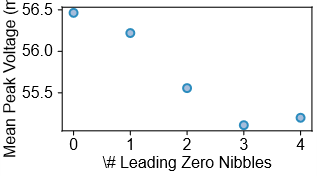}
        \caption{RISC-V}
        \label{fig:vision_go_nibbles}
    \end{subfigure}
    \caption{Sleep-induced power spike vs number of leading zero \textbf{chunks} in the nonce $k$ using \textbf{GoCrypto} ECDSA.}
    \vspace{-0.1in}
    \label{fig:go_nibbles}
\end{figure}

Compared to the scalar multiplication implementations in RustCrypto and BearSSL, GoCrypto utilizes 6-bit chunks during the computation, as explained in Section~\ref{subsec:scale_mul_gocrypto}. The GoCrypto algorithm (with Booth recoding and larger windows) differs enough from the others that generating inputs with more than four leading zero chunks within a reasonable time is difficult. Therefore, the experiments with leading zero chunks for GoCrypto were limited to up to four leading zero chunks, as illustrated in Figure~\ref{fig:go_nibbles}. 

\begin{figure}[h]
    \centering
    \scriptsize
    \begin{subfigure}{0.48\linewidth}
        \centering
        \vspace{-0.04in}
        \includegraphics[width=1\linewidth]{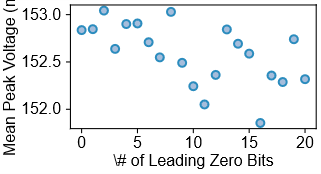}
        \caption{ARM}
        \label{fig:rasp_go_bits}
    \end{subfigure}
    \hfill
    \begin{subfigure}{0.48\linewidth}
        \centering
         \vspace{-0.1in}
         \includegraphics[width=1\linewidth]{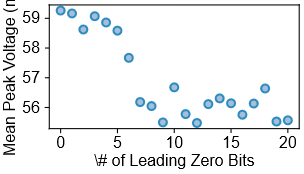}
        \caption{RISC--V}
        \label{fig:vision_go_bits}
    \end{subfigure}
    \caption{Sleep-induced power spike vs number of leading zero \textbf{bits} in the nonce $k$ using \textbf{GoCrypto} ECDSA.}
    \vspace{-0.15in}
    \label{fig:go_bits}
\end{figure}

We also observed that measurements on the Raspberry Pi exhibit substantial fluctuations for the zero-bit experiments, as shown in Figure~\ref{fig:rasp_go_bits}, which makes reliable attack mounting challenging on that platform. In contrast, the VisionFive~2 board (RISC-V) shows clear vulnerabilities across all three libraries.

In summary, we have demonstrated that SleepWalk can be effectively used to extract 20 nonce bits across six configurations (three cryptographic libraries running on two different processor architectures). As described in Section~\ref{subsec:background_ecdsa_hnp}, extraction of 20 nonce bits is enough to recover the secret key using methods like the Hidden Number Problem~\cite{boneh1996hardness}.

\section{Conclusion}
\label{sec:conclusion}

In this paper, we presented a novel power side-channel attack that exploits sleep-induced power spikes to extract ECDSA nonces. Implementations across ARM and RISC-V platforms and multiple libraries (RustCrypto, BearSSL, and GoCrypto) reveal that subtle power variations during CPU context switch induced by sleep expose 20 bits of the nonce. These results confirm that the \texttt{SleepWalk} is a cross-architecture vulnerability that challenges the security of constant-time and deterministic ECDSA implementations.






\vspace{-0.5in}
\begin{IEEEbiography}
[{\includegraphics[width=1in,clip, keepaspectratio]{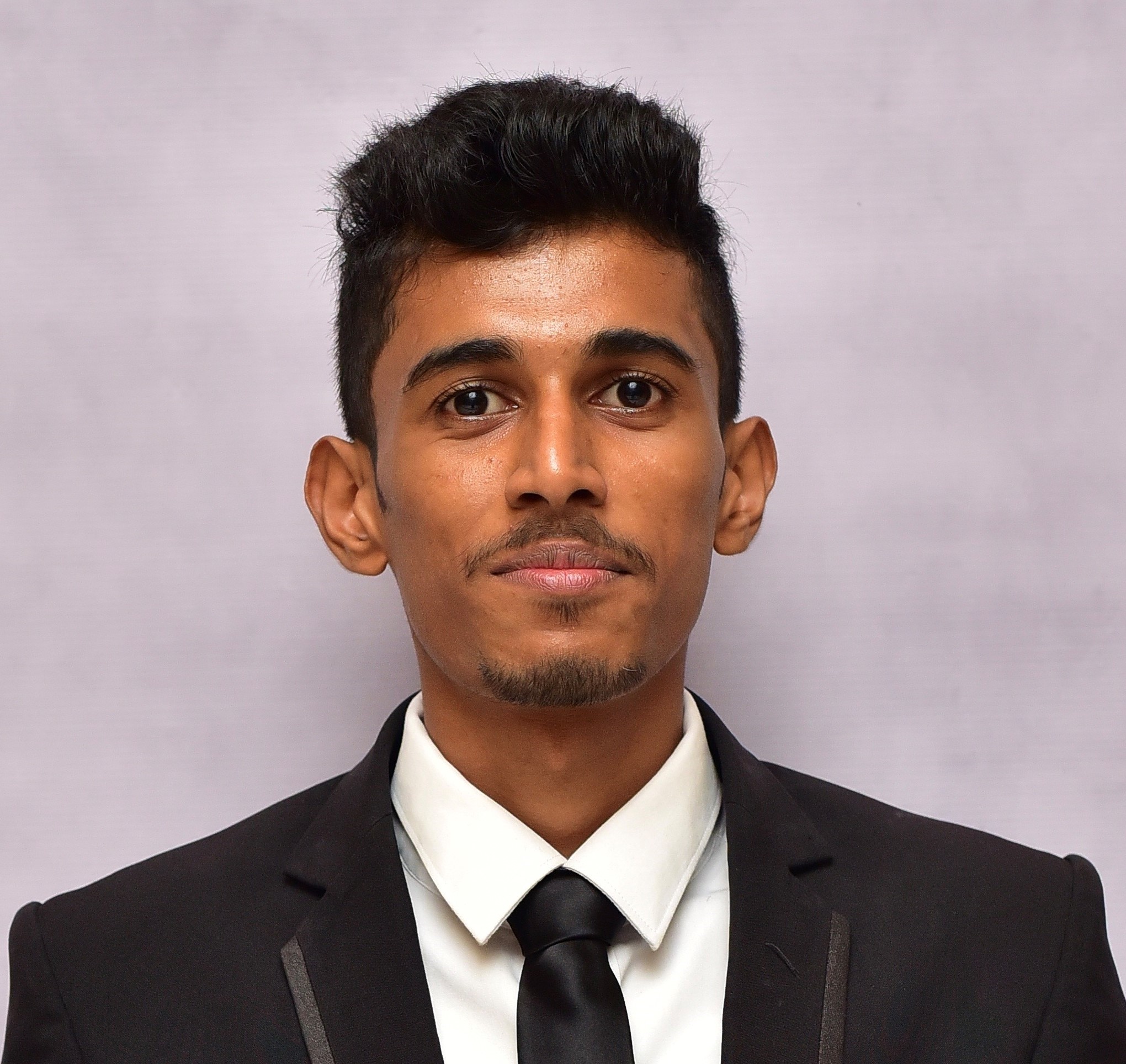}}]{Sahan Sanjaya} is a second-year Ph.D student in the Department of Computer \& Information Science \& Engineering at the University of Florida. In 2022, he completed his B.Sc. in the Department of Electronic and Telecommunication Engineering at the University of Moratuwa, Sri Lanka. His research interests encompass side-channel attacks, hardware security, pre-silicon validation, and post-silicon validation.
\end{IEEEbiography}

\vspace{-0.4 in}
\begin{IEEEbiography}[{\includegraphics[width=1in,clip,keepaspectratio]{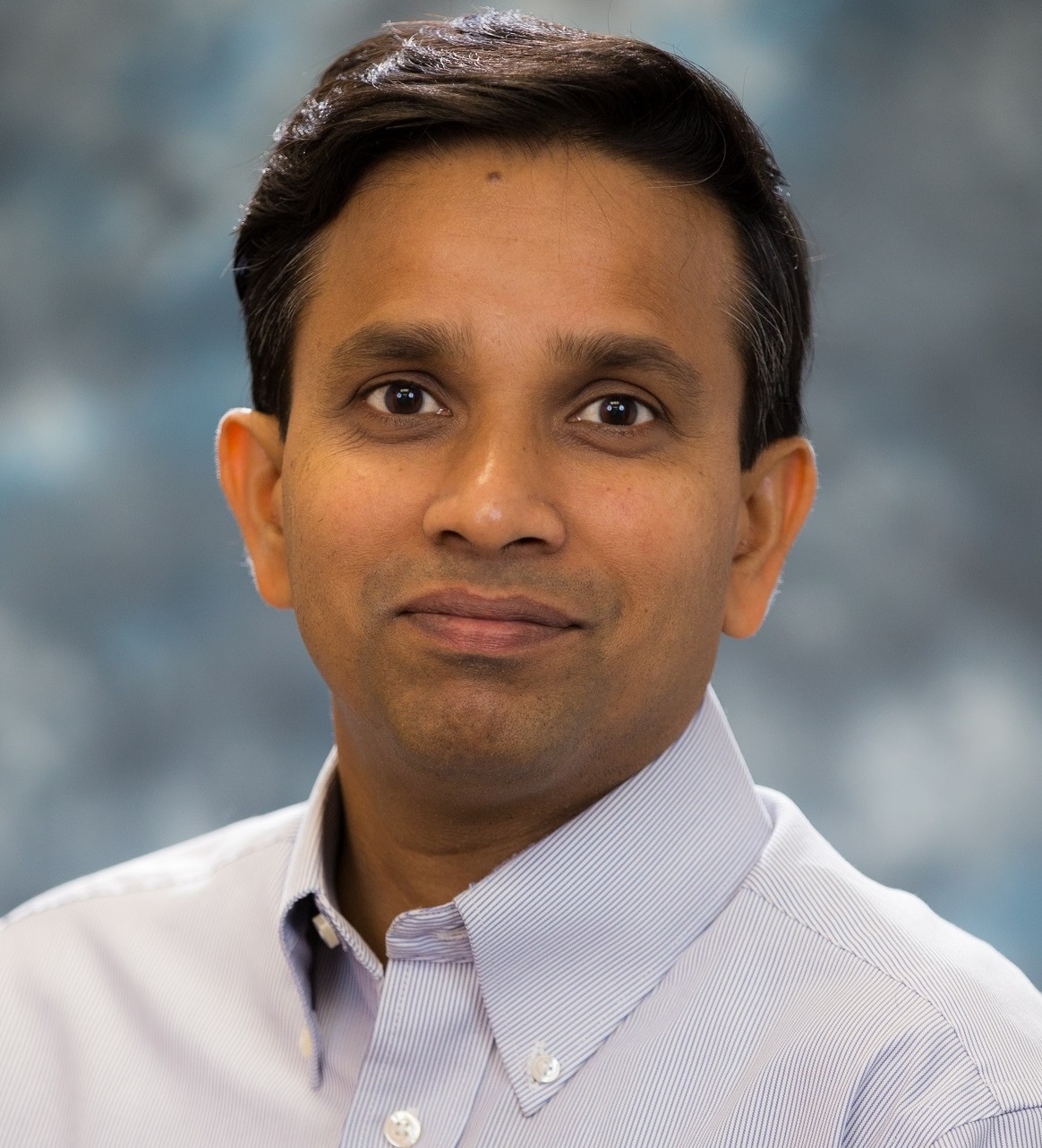}}]{Prabhat Mishra}
is a Professor in the Department of Computer and Information Science and Engineering at the University of Florida. 
His research interests include embedded and cyber-physical systems, hardware security and trust, and energy-aware computing. He currently serves as an Associate Editor of ACM Transactions on Design Automation of Electronic Systems and ACM Transactions on Embedded Computing Systems. He is an IEEE Fellow, an AAAS Fellow, and an ACM Distinguished Scientist.
\end{IEEEbiography}

\newpage
\section{Appendix}\label{sec:appendix}

\begin{listing}[htp]
\caption{Scalar multiplication function in BearSSL}
\label{lst:mul_bearssl}
\begin{lstlisting}[style=vscodeLightC]
static void point_mul_inner(p256_jacobian *R, 
    const p256_affine *W, 
    const unsigned char *k, size_t klen){
    p256_jacobian Q;
    uint32_t qz;
    
    memset(&Q, 0, sizeof Q);
    qz = 1;
    while (klen -- > 0) {
        int i;
        unsigned bk;

        bk = *k++;
        for (i = 0; i < 2; i++) {
            uint32_t bits;
            uint32_t bnz;
            p256_affine T;
            p256_jacobian U;
            uint32_t n;
            int j;
            uint64_t m;

            p256_double(&Q);
            p256_double(&Q);
            p256_double(&Q);
            p256_double(&Q);
            bits = (bk >> 4) & 0x0F;
            bnz  = NEQ(bits, 0);
            /*
             * Lookup point in window. If the bits
             * are 0, we get something invalid, 
             * which is not a problem because we 
             * will use it only if the bits are 
             * non-zero.
             */
            memset(&T, 0, sizeof T);
            for (n = 0; n < 15; n++) {
                m = -(uint64_t)EQ(bits, n + 1);
                T.x[0] |= m & W[n].x[0];
                T.x[1] |= m & W[n].x[1];
                T.x[2] |= m & W[n].x[2];
                T.x[3] |= m & W[n].x[3];
                T.y[0] |= m & W[n].y[0];
                T.y[1] |= m & W[n].y[1];
                T.y[2] |= m & W[n].y[2];
                T.y[3] |= m & W[n].y[3];}
            U = Q;
            p256_add_mixed(&U, &T);
            /*
             * If qz is still 1, then Q was 
             * all-zeros, and this is conserved 
             * through p256_double().
             */
            m = -(uint64_t)(bnz & qz);
            for (j = 0; j < 4; j++) {
                Q.x[j] |= m & T.x[j];
                Q.y[j] |= m & T.y[j];
                Q.z[j] |= m & F256_R[j];}
            CCOPY(bnz & ~qz, &Q, &U, sizeof Q);
            qz &= ~bnz;
            bk <<= 4;
        }
    }
    *R = Q;
}
\end{lstlisting}
\vspace{0.25in}
\end{listing}

\begin{listing}[ht]
\caption{Scalar multiplication function in GoCrypto}
\label{lst:mul_go}
\begin{lstlisting}[style=vscodeLightGo]
func (p *P256Point) p256BaseMult(scalar *p256OrdElement) {
    var t0 p256AffinePoint

     // Extracts the first 6 bits (window) from the scalar.
    wvalue := (scalar[0] << 1) & 0x7f

    // Booth Recoding (boothW6).
    sel, sign := boothW6(uint(wvalue))

    // Precomputed Table (p256Precomputed).
    p256SelectAffine(&t0, &p256Precomputed[0], sel)
    p.x, p.y, p.z = t0.x, t0.y, p256One

    p256NegCond(&p.y, sign)

    index := uint(5)
    zero := sel

    for i := 1; i < 43; i++ {
        // Extracts the next 6 bits from the scalar.
        if index < 192 {
            wvalue = ((scalar[index/64] >> (index % 64)) + (scalar[index/64+1] << (64 - (index % 64)))) & 0x7f
        } else {
            wvalue = (scalar[index/64] >> (index % 64)) & 0x7f
        }

        index += 6
        sel, sign = boothW6(uint(wvalue))
        
        p256SelectAffine(&t0, &p256Precomputed[i], sel)

        p256PointAddAffineAsm(p, p, &t0, sign, sel, zero)
        zero |= sel
    }

    p256MovCond(p, p, NewP256Point(), zero)
}
\end{lstlisting}
\vspace{0.25in}
\end{listing}


\end{document}